\documentclass{rjparticle}
\usepackage{graphicx}

\newcommand{\miktex}{\hbox{Mik\kern-.15em\TeX}}

\title{Neutrino and anti-neutrino energy loss rates due to iron isotopes suitable for core-collapse simulations}
\author[1,2]{Jameel-Un Nabi}
\affil[1]{Faculty of Engineering Sciences, Ghulam Ishaq Khan
Institute, of Engineering Sciences and Technology, Topi 23640,
Swabi, Khyber Pakhtunkhwa, Pakistan \\Email:{\em
jameel@giki.edu.pk}} \affil[2]{Current Address: The Abdus Salam
ICTP, Strada Costiera 11, 34014, Trieste, Italy} \keywords{neutrino
energy loss rates, iron isotopes in stellar matter, pn-QRPA theory,
core-collapse supernovae, core-collapse simulations }
\pacs{97.10.Cv, 26.50.+x, 23.40.Bw, 21.60Jz}

\begin{document}
\maketitle
\begin{abstract}
Accurate estimate of neutrino energy loss rates are needed for the
study of the late stages of the stellar evolution, in particular for
cooling of neutron stars and white dwarfs. The energy spectra of
neutrinos and antineutrinos arriving at the Earth can also provide
useful information on the primary neutrino fluxes as well as
neutrino mixing scenario (it is to be noted that these supernova
neutrinos are emitted after the supernova explosion which is a much
later stage of stellar evolution than that considered in this
paper). Recently an improved microscopic calculation of
weak-interaction mediated rates for iron isotopes was introduced
using the proton-neutron quasiparticle random phase approximation
(pn-QRPA) theory. Here I present for the first time the fine-grid
calculation of the neutrino and anti-neutrino energy loss rates due
to $^{54,55,56}$Fe in stellar matter. In the core of massive stars
isotopes of iron, $^{54,55,56}$Fe, are considered to be key players
in decreasing the electron-to-baryon ratio ($Y_{e}$) mainly via
electron capture on these nuclide. Core-collapse simulators may find
this calculation suitable for interpolation purposes and for
necessary incorporation in the stellar evolution codes. The
calculated cooling rates are also compared with previous
calculations.
\end{abstract}
\section{ Introduction}

Neutrino processes and electron capture play pivotal roles in
core-collapse supernova explosions. The discovery of neutrino burst
from SN1987A by the Kamiokande II group \cite{Hir87} and IMB group
\cite{Bio87} energized the research on neutrino astrophysics.
According to the stellar evolution theory and the observation data,
the burst of neutrino is the first signal from the supernova
explosion. The core of a massive star implodes in a time period of
0.5-1 s and around 10$^{53}$ ergs of binding energy of gravity is
released, of which only 1$\%$ is transferred by neutrinos to eject
the envelope of the star. Understanding neutrino interactions in the
circumstance of a supernova is pivotal to a better understanding of
the explosion mechanism.

During the first tens of seconds of the proto-neutron star's
evolution, neutrinos diffuse outward and escape from the star
thereby lowering the entropy and lepton abundance of the stellar
matter. This scenario does not necessarily hold at extremely high
densities and temperatures (this would be the case for stellar
collapse where dynamical time scales become shorter than the
neutrino transport time scales) where neutrinos can become trapped
in the so-called neutrinospheres mainly due to elastic scattering
with nuclei. During the late stages of stellar evolution a star
mainly looses energy through neutrinos. White dwarfs and supernovae
(which are the endpoints for stars of varying masses) have both
cooling rates largely dominated by neutrino production. Prior to
stellar collapse one requires an accurate determination of neutrino
energy loss rates in order to perform a careful study of the final
branches of star evolutionary tracks. A change in the cooling rates
particularly at the very last stages of massive star evolution could
affect the evolutionary time scale and the iron core configuration
at the onset of the explosion \cite{Esp03}. The electron capture
rates and the accompanying neutrino energy loss rates are also
required in determining the equation of state. Reliable and
microscopic calculations of neutrino cooling rates and capture rates
can contribute effectively in the final outcome of these simulations
on world's fastest supercomputers.

The first-ever extensive calculation of stellar weak rates including
the capture rates, neutrino energy loss rates and decay rates for a
wide density and temperature domain was performed by Fuller, Fowler,
and Newman (FFN) \cite{Ful80}. The calculation was done for 226
nuclei in the mass range $21 \leq A \leq 60$. The authors stressed
on the importance of the Gamow-Teller (GT) giant resonance strength
in the capture of the electron and estimated the GT centroids using
zeroth-order ($0\hbar\omega$ ) shell model. The measured data from
various (p,n) and (n,p) experiments later revealed the misplacement
of the GT centroid adopted in the parameterizations of FFN. Since
then theoretical efforts were concentrated on the microscopic
calculations of weak-interaction mediated rates of iron-regime
nuclide.  The proton-neutron quasiparticle random phase
approximation (pn-QRPA) theory  (e.g. \cite{Nab04}) and the
large-scale shell model (LSSM)(e.g. \cite{Lan00}) were used
extensively and with relative success for the microscopic
calculation of stellar weak rates.

Proton-neutron quasi particle random phase approximation (pn-QRPA)
theory and shell model are extensively used for the calculations of
capture rates in the stellar environment. Shell model lays more
emphasis on interaction of nucleons as compared to correlations
whereas pn-QRPA puts more weight on correlations. One big advantage
of using pn-QRPA theory is that it gives us the liberty of
performing calculations in a luxurious model space (up to
$7\hbar\omega$). The pn-QRPA method considers the residual
correlations among the nucleons via one particle one hole (1p-1h)
excitations in a large model spaces. The prevailing temperature of
the stellar matter is of the order of a few hundred kilo-electron
volts to a few million-electron volts and GT$_{\pm}$ transitions
occur not only from nuclear ground state, but also from excited
states. As experimental information about excited state strength
functions seems inaccessible, Aufderheide \cite{Auf91} stressed much
earlier the need to probe these strength functions theoretically.
Today the pn-QRPA theory calculates the GT$_{\pm}$ strength
distribution of \textit{all} excited states of parent nucleus in a
microscopic fashion and this feature of the pn-QRPA model greatly
enhances the reliability of the calculated rates in stellar matter.
In other words the pn-QRPA model allows a microscopic state-by-state
calculation of all stellar weak rates and the Brink hypothesis is
not assumed in this model. Brink hypothesis states that GT strength
distribution on excited states is \textit{identical} to that from
ground state, shifted \textit{only} by the excitation energy of the
state. Recent calculations have pointed towards the fact that Brink
hypothesis is not a safe approximation to use for calculation of
stellar weak-interaction rates \cite{Nab08, Nab09, Nab10, Nab10a,
Nab11}.

During the presupernova evolution of massive stars, the isotopes of
iron, $^{54,55,56}$Fe, are advocated to play a key role inside the
cores primarily decreasing the electron-to-baryon ratio ($Y_{e}$)
mainly via electron capture processes thereby reducing the pressure
support. The neutrinos and antineutrinos produced, as a result of
these weak-interaction reactions, are transparent to the stellar
matter and assist in cooling the core thereby reducing the entropy.
The structure of the presupernova star is altered both by the
changes in $Y_{e}$ and the entropy of the core material. Calculation
of weak-interaction rates of isotopes of iron, $^{54,55,56}$Fe, in
stellar matter, were recently calculated using the pn-QRPA model
\cite{Nab09}. Later on the detailed analysis of neutrino cooling
rates due to these isotopes of iron relevant to the presupernova
evolution of massive stars were presented \cite{Nab11a}. There the
author reported that during the presupernova evolution of massive
stars, from oxygen shell burning till around end of convective core
silicon burning phases, the pn-QRPA calculated neutrino cooling
rates due to $^{54}$Fe were three to four times bigger than the
large-scale shell model (LSSM) results \cite{Lan00}. For other
phases of presupernova evolution, at times the two rates were in
very good comparison, while at other instances the reported neutrino
cooling rates were slightly bigger or smaller than the corresponding
LSSM rates. The temporal variation of $Y_{e}$ depends on how fine
the calculated weak rates changes with time. Due to the rather small
variation in calculated rates, a fine-grid calculation of stellar
neutrino and anti-neutrino energy loss rates due to $^{54,55,56}$Fe
was desired by the core-collapse simulators for more accurate
outcomes.  Due to the extreme conditions prevailing in the cores of
massive stars, interpolation of calculated rates within large
intervals of temperature-density points posed some uncertainty in
the values of cooling rates for collapse simulators. For the first
time a fine-grid calculation of neutrino and anti-neutrino energy
loss rates in stellar matter due to isotopes of iron  is being
presented in this paper. The calculation is done using the pn-QRPA
model and is presented on a detailed temperature and density grid
pertinent to presupernova and supernova environment and should prove
more suitable for running on simulation codes. It is to be noted
that throughout this paper (anti)neutrino energy loss rates and
(anti)neutrino cooling rates are meant as the same physical
phenomena and the two terms are used interchangeably.

Section 2 briefly discusses the formalism of the pn-QRPA model and
presents some of the calculated results. Comparison with earlier
calculations is also included in this section. I summarize the main
conclusions in Section 3. The fine-grid calculation of neutrino and
anti-neutrino energy loss rates due to $^{54,55,56}$Fe is presented
in a tabular form towards the end of this manuscript.

\section{ Calculation and Results}
The neutrino and anti-neutrino energy loss rates of a transition
from the $ith$ state of the parent to the $jth$ state of the
daughter nucleus is given by
\begin{equation}
\lambda ^{^{\nu(\bar{\nu})} } _{ij} =\left[\frac{\ln 2}{D}
\right]\left[f_{ij} (T,\rho ,E_{f} )\right]\left[B(F)_{ij}
+\left({\raise0.7ex\hbox{$ g_{A}  $}\!\mathord{\left/ {\vphantom
{g_{A}  g_{V} }} \right.
\kern-\nulldelimiterspace}\!\lower0.7ex\hbox{$ g_{V}  $}}
\right)^{2} B(GT)_{ij} \right].
\end{equation}
The value of D was taken to be 6295s \cite{Yos88}. $B_{ij}'s$ are
the sum of reduced transition probabilities of the Fermi B(F) and GT
transitions B(GT). Details of the calculations of phase space
integrals and reduced transition probabilities in the pn-QRPA model
can be found in Refs. \cite{Nab11a, Nab10b}.

The total (anti)neutrino energy loss rate per unit time per nucleus
was then calculated using
\begin{equation}
\lambda^{\nu(\bar{\nu})} =\sum _{ij}P_{i} \lambda
_{ij}^{\nu(\bar{\nu})}.
\end{equation}
The summation over all initial and final states was carried out
until satisfactory convergence in the rate calculations was
achieved. Here $P_{i}$ is the probability of occupation of parent
excited states and follows the normal Boltzmann distribution. The
pn-QRPA theory allows a microscopic state-by-state calculation of
both sums present in Eq. (2). As discussed earlier this feature of
the pn-QRPA model greatly increases the reliability of the
calculated rates in stellar matter where there exists a finite
probability of occupation of excited states.

The deformation parameter was recently being argued as an important
parameter for QRPA calculations at par with the pairing parameter by
Stetcu and Johnson \cite{Ste04}. As such rather than using
deformations from some theoretical mass model (as used in earlier
calculations of pn-QRPA rates e.g. \cite{Nab04, Nab99}) the
experimentally adopted value of the deformation parameters for
$^{54,56}$Fe, extracted by relating the measured energy of the first
$2^{+}$ excited state with the quadrupole deformation, was taken
from Raman et al. \cite{Ram87}. For the case of $^{55}$Fe (where
such measurement lacks) the deformation of the nucleus was
calculated as
\begin{equation}
\delta = \frac{125(Q_{2})}{1.44 (Z) (A)^{2/3}},
\end{equation}
where $Z$ and $A$ are the atomic and mass numbers, respectively, and
$Q_{2}$ is the electric quadrupole moment taken from Ref.
\cite{Moe81}. Q-values were taken from the recent mass compilation
of Audi et al. \cite{Aud03}.

The incorporation of measured deformations for $^{54,56}$Fe and a
smart choice of strength parameters led to an improvement of the
calculated GT$_{\pm}$ distributions compared to the measured ones
\cite{Nab09}. It was shown in Table 1 of Ref. \cite{Nab09} that the
present pn-QRPA calculated GT$_{\pm}$ centroids and total
$S_{\beta^{\pm}}$ strengths were in good agreement with available
data for the even-even isotopes of iron. The table also showed the
marked improvement in the reported pn-QRPA calculation over the
previous one \cite{Nab04}.

In order to further increase the reliability of the calculated
capture rates experimental data were incorporated in the calculation
wherever possible. In addition to the incorporation of the
experimentally adopted value of the deformation parameter, the
calculated excitation energies (along with their log $ft$ values)
were replaced with an experimental one when they were within 0.5 MeV
of each other. Missing measured states were inserted and inverse and
mirror transitions were also taken into account. No theoretical
levels were replaced with the experimental ones beyond the
excitation energy for which experimental compilations had no
definite spin and/or parity. A state-by-state calculation of
GT$_{\pm}$ strength was performed for a total of 246 parent excited
states in $^{54}$Fe, 297 states for $^{55}$Fe and 266 states for
$^{56}$Fe. For each parent excited state, transitions were
calculated to 150 daughter excited states. The band widths of energy
states were chosen according to the density of states to cover an
excitation energy of (15-20) MeV in parent and daughter nuclei. The
summation in Eq. (2) was done to ensure satisfactory convergence.
The use of a separable interaction assisted in the incorporation of
a luxurious model space of up to 7 major oscillator shells which in
turn made possible to consider these many excited states both in
parent and daughter nucleus.

The calculation of neutrino and anti-neutrino energy loss rates due
to $^{54,55,56}$Fe was presented earlier in Ref. \cite{Nab11a}. In
this paper I would like to present a graphical comparison of these
energy loss rates with previous calculations for stellar densities
$\rho Y_{e} [gcm^{-3}] =10^{6}, 10^{7}, 10^{8}$. The temperature
ranges from $T_{9} [K] = 1$ to $T_{9} [K] = 30$. This
density-temperature domain is particularly important for the
presupernova evolution of massive stars (see Ref. \cite{Nab11a} for
details). The pn-QRPA calculation is compared with the results of
large scale shell model (LSSM)  \cite{Lan00} and the pioneering
calculation of FFN \cite{Ful80}.

\begin{figure}[htbp]
\begin{center}
\includegraphics[width=0.8\textwidth]{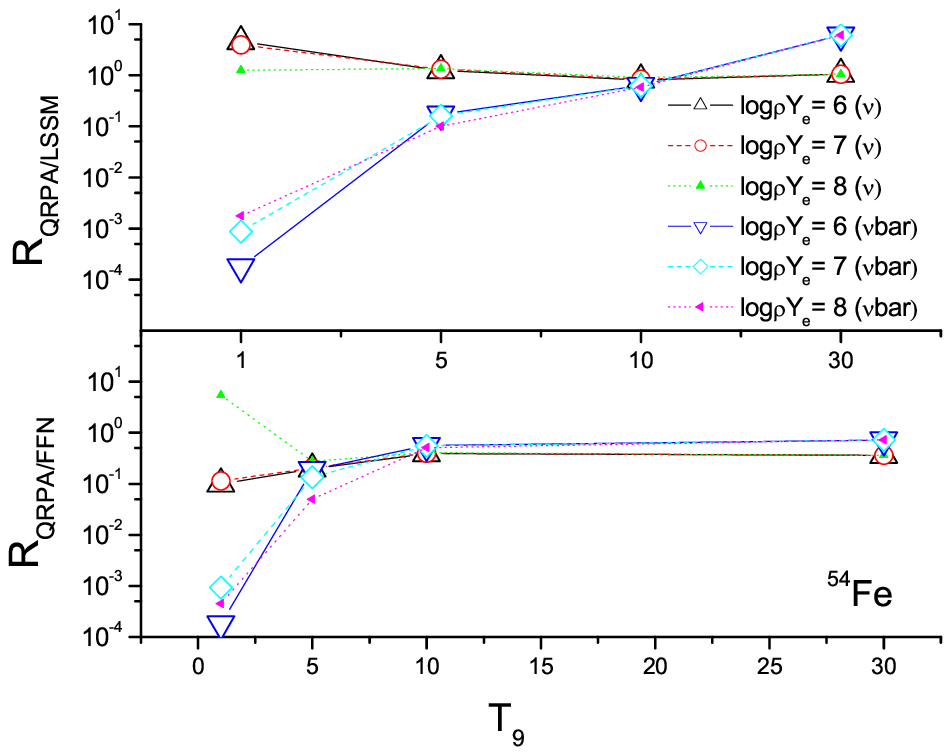}
\caption{(Color online) Ratios of reported neutrino ($\nu$) and
anti-neutrino ($\nu$bar) cooling rates due to $^{54}$Fe to those
calculated using LSSM \cite{Lan00} (upper panel) and FFN
\cite{Ful80} (lower panel) as function of stellar temperatures and
densities. T$_{9}$ gives the stellar temperature in units of
$10^{9}$ K. In the legend, $log \rho Y_{e}$ gives the log to base 10
of stellar density in units of $gcm^{-3}$, $\nu$ and $\nu$bar stand
for neutrino and anti-neutrino cooling rate ratios, respectively.}
\label{fig1}
\end{center}
\end{figure}

The pn-QRPA calculated neutrino and anti-neutrino energy loss rates
due to $^{54}$Fe is compared against LSSM and FFN calculations in
Fig.~\ref{fig1}. The upper panel displays the ratio of calculated
rates to the LSSM rates, $R_{QRPA/LSSM}$, while the lower panel
shows a similar comparison with the FFN calculation, $R_{QRPA/FFN}$.
Both neutrino and anti-neutrino energy loss rate comparisons are
presented in each panel. It can be seen that at low temperatures and
densities the calculated neutrino energy loss rates is around five
times bigger than those calculated by LSSM. During the oxygen shell
burning and silicon core burning phases of massive stars the pn-QRPA
energy loss rates are around 3--4 times bigger than LSMM rates. At
high temperatures and densities the two calculations are in very
good agreement. It is pertinent to mention again that the neutrino
energy loss rates contain contributions due to electron capture and
positron decay on iron isotopes. Since the electron capture rates
are orders of magnitude bigger than the corresponding positron decay
rates \cite{Nab09} the neutrino energy loss rate comparison is
dictated by the corresponding comparison between electron capture
rates. The calculated neutrino energy loss rates due to $^{54}$Fe
are up to an order of magnitude smaller  as compared to FFN rates at
$\rho Y_{e} [gcm^{-3}] =10^{6}, 10^{7}$ and low temperatures (lower
panel of Fig.~\ref{fig1}). The comparison improves at higher
temperatures (where the FFN rates are roughly a factor three
bigger). It is to be noted that FFN neglected the quenching of the
total GT strength in their rate calculation. The calculated
antineutrino energy loss rates contain contributions due to positron
capture and $\beta$-decay on iron isotopes. Both these contributions
are relatively very small as compared to the corresponding electron
capture rates. Correspondingly the anti-neutrino energy loss rates
are very small in magnitude. These small numbers are fragile
functions of the available phase space and can change appreciably by
a mere change of 0.5 MeV in phase space calculations and are more
reflective of the uncertainties in calculation of the energy
eigenvalues (for both parent and daughter states). At $T_{9} [K] =
1$ and $\rho Y_{e} [gcm^{-3}] =10^{6}$, the pn-QRPA anti-neutrino
energy loss rates are up to four orders of magnitude smaller than
the corresponding LSSM rates (upper panel). At high temperatures the
two calculations are in reasonable agreement with the pn-QRPA rates
bigger by a factor of six. It can be seen that the anti-neutrino
energy loss rates are more dependent on the stellar temperatures.
The comparison with FFN calculations follows a similar trend (lower
panel). However the FFN rates are bigger at all temperature-density
domain.

\begin{figure}[htbp]
\begin{center}
\includegraphics[width=0.8\textwidth]{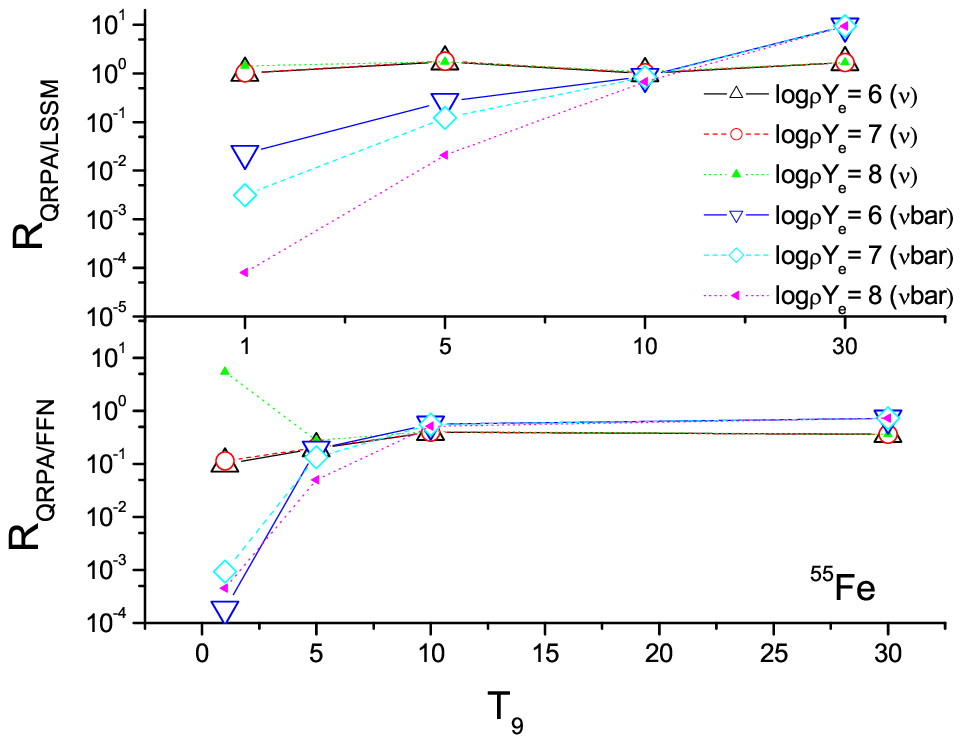}
\caption{(Color online) Same as Fig.~\ref{fig1} but for cooling
ratios due to  $^{55}$Fe.} \label{fig2}
\end{center}
\end{figure}

Fig.~\ref{fig2} shows how the calculated  energy loss rates compare
with previous calculations due to $^{55}$Fe. The upper panel of
Fig.~\ref{fig2} shows that LSSM and pn-QRPA calculations agree very
well for all important temperature-density domain. The FFN rates are
bigger by an order of magnitude at low temperatures and stellar
densities in the range $\rho Y_{e} [gcm^{-3}] =10^{6}, 10^{7}$
(lower panel). The probability of occupation of high-lying excited
states increases with stellar temperature and FFN did not take into
effect the process of particle emission from excited states and
accordingly their parent excitation energies extended well beyond
the particle decay channel. The anti-neutrino energy loss rated due
to $^{55}$Fe are around four orders of magnitude smaller than LSSM
rates at $T_{9} [K] = 1$ and $\rho Y_{e} [gcm^{-3}] =10^{8}$. At
$T_{9} [K] = 30$ the pn-QRPA rates are around an order of magnitude
bigger (upper panel).  The FFN anti-neutrino energy loss rates are
up to four orders of magnitude bigger than pn-QRPA rates at low
temperatures. The comparison is reasonably well at high temperatures
(lower panel).

\begin{figure}[htbp]
\begin{center}
\includegraphics[width=0.8\textwidth]{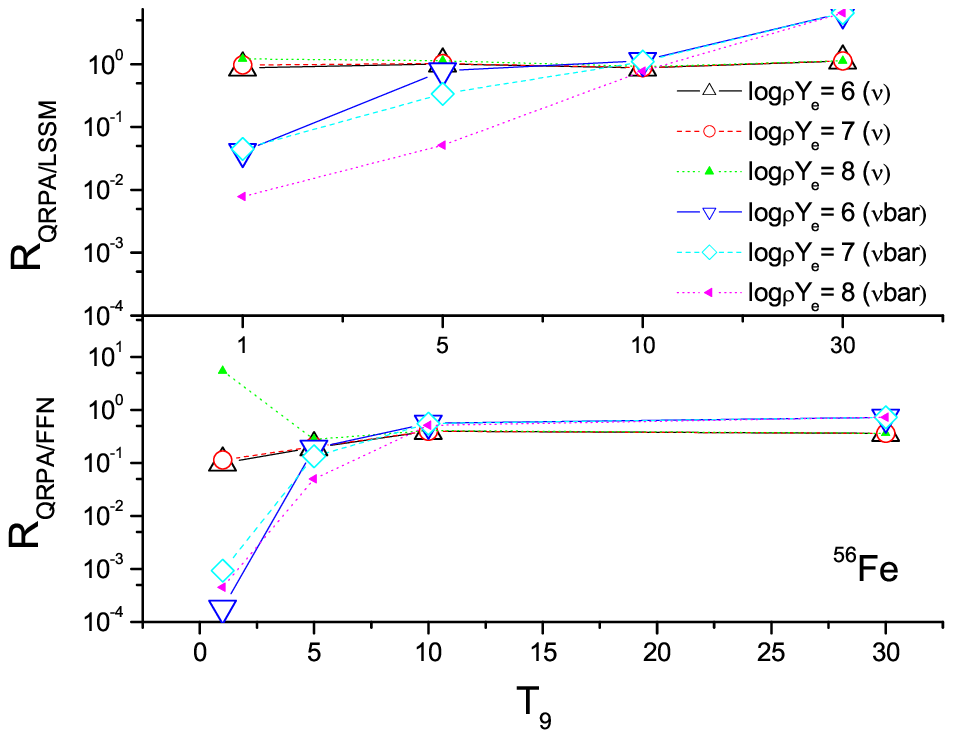}
\caption{(Color online) Same as Fig.~\ref{fig1} but for cooling
ratios due to $^{56}$Fe.} \label{fig3}
\end{center}
\end{figure}

Comparison of the pn-QRPA calculated neutrino energy loss rates due
to $^{56}$Fe with LSSM is again reasonably well (Fig.~\ref{fig3})
for the astrophysically important density-temperature domain. The
electron capture rates on $^{56}$Fe are very important for the
pre-supernova phase of massive stars. FFN rates (lower panel) are
again bigger for reasons already mentioned. Regarding the comparison
of anti-neutrino energy loss rates the upper panel shows that at low
stellar temperatures the LSSM rates are bigger by 1--2 orders of
magnitude. The comparison is fairly well at $T_{9} [K] = 10$ while
at still higher temperatures the pn-QRPA rates are bigger by a
factor of 7 for all density range shown in Fig.~\ref{fig3}. FFN
anti-neutrino energy loss rates are around 4  orders of magnitude
bigger at low temperatures. The comparison improves as the stellar
temperature increases. For a detailed discussion on the possible
reasons for these differences, the reader is referred to Ref.
\cite{Nab11a}.

The fine-grid calculation of pn-QRPA calculated energy loss rates
due to $^{54,55,56}$Fe is presented in Table~1. The calculated rates
are tabulated in logarithmic (to base 10) scale. The first column
gives stellar densities, log($\rho Y_{e}$), in units of $g cm^{-3}$,
where $\rho$ is the baryon density and $Y_{e}$ is the ratio of the
electron number to the baryon number. Stellar temperatures ($T_{9}$)
are given in $10^{9}$ K. Stated also are the values of the Fermi
energy of electrons in units of MeV.
$\lambda^{\nu}$($\lambda^{\bar{\nu}}$) are the
neutrino(anti-neutrino) energy loss rates in units of  $MeV.s^{-1}$.
It is to be noted that Table~1 only shows the calculated rates at a
selected density of $\rho Y_{e} [gcm^{-3}] =10^{6.5}$ and $10^{7}$.
The complete table is not presented here to save space. Interested
readers may find a compete set of table covering stellar density in
the range $\rho Y_{e} [gcm^{-3}] =10 - 10^{11}$ as an addendum  to
the on-line version of this paper. Core-collapse simulators may find
Table~1 useful for interpolation purposes and check the consequences
of incorporating the reported energy cooling rates due to isotopes
of iron in their simulation codes. The ASCII file of Table~1 is also
available and can be received from the author upon request.

\section{ Summary}
In order to understand the supernova explosion mechanism
international collaborations of astronomers and physicists are being
sought. Weak interaction mediated rates are key nuclear physics
input to simulation codes and a reliable and  microscopic
calculation of these rates (both from ground-state \textit{and}
excited states) is desirable. The pn-QRPA model has a good track
record of calculation of weak interaction rates both in terrestrial
and stellar domains. The model has access to a huge model space
making it possible to calculate weak rates for arbitrarily heavy
system of nucleons. Further the model gets rid of the Brink
hypothesis and calculates a \textit{state-by-state} calculation of
stellar capture rates which greatly increases the reliability of
calculated rates. Incorporation of experimental deformation lead to
a much improved version of this calculation. The model was used
recently to calculate weak-interaction mediated rates on iron
isotopes, $^{54,55,56}$Fe \cite{Nab09}.

The main idea of reporting this work is to present a fine-grid
calculation of neutrino and anti-neutrino energy loss rates due to
$^{54,55,56}$Fe in stellar matter. Table 1 shows the result. As
mentioned earlier the complete table can be found in the addendum of
the on-line version of this paper. This table can be of great
utility for core-collapse simulators and is more convenient for
interpolation purposes. The calculation was also compared against
LSSM and FFN calculations. The pn-QRPA neutrino energy loss rates
due to $^{54}$Fe is around five times bigger than LSSM rates and
favor cooler stellar cores. The two calculations are in good
agreement for $^{55,56}$Fe. FFN rates on the average are around an
order of magnitude bigger.

\begin{acknowledgement}
The author would like to acknowledge the local hospitality provided
by the Abdus Salam ICTP, Trieste, where part of this project was
completed. The author wishes to acknowledge the support of research
grant provided by the Higher Education Commission, Pakistan  through
the HEC Project No. 20-1283.
\end{acknowledgement}

\newpage

\scriptsize\clearpage \textbf{Table 1:} Calculated neutrino and
anti-neutrino energy loss rates due to $^{54,55,56}$Fe on a
fine-grid temperature scale.   The calculated rates are tabulated in
logarithmic (to base 10) scale. The rates are tabulated for a
selected density of $log \rho Y_{e} = 6.5$ and $7$.  For units see
text. For complete table see the addendum in the on-line version of
this paper.\label{tab.1}
\begin{center}
\scriptsize\begin{tabular}{|ccc|cccccc|} \hline\hline $log\rho
Y_{e}$ & $T_{9}$ & $E_{f}$ & \multicolumn{2}{|c|}{$^{54}$Fe}&
\multicolumn{2}{|c|}{$^{55}$Fe}
& \multicolumn{2}{|c|}{$^{56}$Fe} \\
\cline{4-9} & & & \multicolumn{1}{c}{$\lambda_{\nu}$} &
\multicolumn{1}{|c|}{$\lambda_{\bar{\nu}}$} &
\multicolumn{1}{c}{$\lambda_{\nu}$} &
\multicolumn{1}{|c|}{$\lambda_{\bar{\nu}}$} &
\multicolumn{1}{c}{$\lambda_{\nu}$} &
\multicolumn{1}{|c|}{$\lambda_{\bar{\nu}}$} \\ \hline
   6.5 &   0.50 &   0.905 &    -18.321 &    -96.473 &     -5.976 &    -48.612 &    -39.992 &    -59.560   \\
   6.5 &   1.00 &   0.880 &    -11.149 &    -51.087 &     -5.851 &    -27.623 &    -21.887 &    -32.715   \\
   6.5 &   1.50 &   0.837 &     -8.661 &    -34.623 &     -5.653 &    -19.939 &    -15.468 &    -22.696   \\
   6.5 &   2.00 &   0.777 &     -7.303 &    -26.161 &     -5.320 &    -15.620 &    -12.113 &    -17.365   \\
   6.5 &   2.50 &   0.701 &     -6.373 &    -20.942 &     -4.907 &    -12.889 &    -10.025 &    -14.035   \\
   6.5 &   3.00 &   0.612 &     -5.656 &    -17.358 &     -4.511 &    -10.983 &     -8.584 &    -11.727   \\
   6.5 &   3.50 &   0.517 &     -5.067 &    -14.726 &     -4.155 &     -9.554 &     -7.513 &    -10.013   \\
   6.5 &   4.00 &   0.424 &     -4.561 &    -12.710 &     -3.830 &     -8.429 &     -6.667 &     -8.684   \\
   6.5 &   4.50 &   0.343 &     -4.110 &    -11.120 &     -3.524 &     -7.514 &     -5.965 &     -7.622   \\
   6.5 &   5.00 &   0.277 &     -3.698 &     -9.837 &     -3.231 &     -6.752 &     -5.360 &     -6.755   \\
   6.5 &   5.50 &   0.226 &     -3.320 &     -8.779 &     -2.952 &     -6.104 &     -4.829 &     -6.029   \\
   6.5 &   6.00 &   0.187 &     -2.971 &     -7.888 &     -2.688 &     -5.541 &     -4.358 &     -5.410   \\
   6.5 &   6.50 &   0.157 &     -2.650 &     -7.126 &     -2.439 &     -5.045 &     -3.934 &     -4.873   \\
   6.5 &   7.00 &   0.134 &     -2.353 &     -6.464 &     -2.204 &     -4.601 &     -3.551 &     -4.400   \\
   6.5 &   7.50 &   0.115 &     -2.077 &     -5.881 &     -1.983 &     -4.199 &     -3.202 &     -3.978   \\
   6.5 &   8.00 &   0.100 &     -1.820 &     -5.363 &     -1.773 &     -3.833 &     -2.881 &     -3.598   \\
   6.5 &   8.50 &   0.088 &     -1.581 &     -4.898 &     -1.574 &     -3.496 &     -2.586 &     -3.253   \\
   6.5 &   9.00 &   0.078 &     -1.356 &     -4.477 &     -1.384 &     -3.183 &     -2.312 &     -2.937   \\
   6.5 &   9.50 &   0.069 &     -1.146 &     -4.094 &     -1.202 &     -2.892 &     -2.057 &     -2.645   \\
   6.5 &  10.00 &   0.062 &     -0.947 &     -3.743 &     -1.028 &     -2.620 &     -1.818 &     -2.375   \\
   6.5 &  15.00 &   0.027 &      0.556 &     -1.328 &      0.426 &     -0.601 &     -0.039 &     -0.430   \\
   6.5 &  20.00 &   0.015 &      1.530 &      0.081 &      1.481 &      0.675 &      1.094 &      0.776   \\
   6.5 &  25.00 &   0.010 &      2.235 &      1.051 &      2.263 &      1.569 &      1.896 &      1.628   \\
   6.5 &  30.00 &   0.007 &      2.784 &      1.785 &      2.866 &      2.242 &      2.509 &      2.281   \\
   7.0 &   0.50 &   1.217 &    -17.541 &    -99.304 &     -5.049 &    -51.753 &    -36.865 &    -62.102   \\
   7.0 &   1.00 &   1.200 &    -10.387 &    -52.135 &     -4.985 &    -29.221 &    -20.275 &    -33.771   \\
   7.0 &   1.50 &   1.173 &     -7.919 &    -35.664 &     -4.870 &    -21.054 &    -14.341 &    -23.569   \\
   7.0 &   2.00 &   1.133 &     -6.587 &    -27.040 &     -4.628 &    -16.515 &    -11.215 &    -18.157   \\
   7.0 &   2.50 &   1.083 &     -5.695 &    -21.706 &     -4.280 &    -13.657 &     -9.256 &    -14.751   \\
   7.0 &   3.00 &   1.021 &     -5.019 &    -18.043 &     -3.922 &    -11.670 &     -7.898 &    -12.381   \\
   7.0 &   3.50 &   0.950 &     -4.475 &    -15.349 &     -3.600 &    -10.177 &     -6.890 &    -10.614   \\
   7.0 &   4.00 &   0.871 &     -4.021 &    -13.272 &     -3.316 &     -8.991 &     -6.106 &     -9.230   \\
   7.0 &   4.50 &   0.785 &     -3.630 &    -11.616 &     -3.062 &     -8.010 &     -5.470 &     -8.105   \\
   7.0 &   5.00 &   0.698 &     -3.285 &    -10.262 &     -2.831 &     -7.177 &     -4.937 &     -7.169   \\
   7.0 &   5.50 &   0.613 &     -2.974 &     -9.133 &     -2.615 &     -6.458 &     -4.476 &     -6.376   \\
   7.0 &   6.00 &   0.534 &     -2.687 &     -8.180 &     -2.409 &     -5.832 &     -4.067 &     -5.696   \\
   7.0 &   6.50 &   0.465 &     -2.417 &     -7.364 &     -2.210 &     -5.283 &     -3.697 &     -5.107   \\
   7.0 &   7.00 &   0.404 &     -2.162 &     -6.659 &     -2.016 &     -4.795 &     -3.357 &     -4.591   \\
   7.0 &   7.50 &   0.353 &     -1.921 &     -6.041 &     -1.828 &     -4.359 &     -3.043 &     -4.135   \\
   7.0 &   8.00 &   0.310 &     -1.691 &     -5.495 &     -1.645 &     -3.965 &     -2.750 &     -3.728   \\
   7.0 &   8.50 &   0.274 &     -1.473 &     -5.008 &     -1.467 &     -3.606 &     -2.476 &     -3.361   \\
   7.0 &   9.00 &   0.244 &     -1.266 &     -4.570 &     -1.294 &     -3.276 &     -2.220 &     -3.028   \\
   7.0 &   9.50 &   0.218 &     -1.069 &     -4.173 &     -1.126 &     -2.971 &     -1.978 &     -2.723   \\
   7.0 &  10.00 &   0.196 &     -0.881 &     -3.810 &     -0.962 &     -2.688 &     -1.751 &     -2.442   \\
   7.0 &  15.00 &   0.085 &      0.575 &     -1.347 &      0.445 &     -0.621 &     -0.020 &     -0.449   \\
   7.0 &  20.00 &   0.047 &      1.538 &      0.073 &      1.489 &      0.666 &      1.102 &      0.768   \\
   7.0 &  25.00 &   0.030 &      2.239 &      1.047 &      2.267 &      1.564 &      1.900 &      1.624   \\
   7.0 &  30.00 &   0.021 &      2.786 &      1.783 &      2.868 &      2.239 &      2.511 &      2.279   \\
\end{tabular}
\end{center}

\end{document}